\documentclass[%
 reprint,
 amsmath,amssymb,
 aps,
]{revtex4-2}

\usepackage{graphicx}%
\usepackage{dcolumn}%
\usepackage{bm}%

\begin{document}

\title{Temporal equivalent of the Sommerfeld half-plane problem}%

\author{Asma Fallah\textsuperscript{1}}
\author{Shixiong Yin\textsuperscript{2,}\textsuperscript{3}}

\author{Andrea Alù\textsuperscript{2,}\textsuperscript{3,}\textsuperscript{4}}
\email{aalu@gc.cuny.edu}

 \author{Nader Engheta\textsuperscript{1}}%
 \email{engheta@seas.upenn.edu}

\affiliation{%
 \textsuperscript{1} University of Pennsylvania, Department of Electrical and Systems Engineering, Philadelphia, Pennsylvania 19104, USA
}%

\affiliation{%
\textsuperscript{2} Photonics Initiative, Advanced Science Research Center, City University of New York, New York, New York 10031, USA}%
\affiliation{%
\textsuperscript{3} Department of Electrical Engineering, City College of The City University of New York, New York, New York 10031, USA}%
\affiliation{%
\textsuperscript{4} Physics Program, Graduate Center, City University of New York, New York, New York 10016, USA}%

\date{}%

\begin{abstract}

Sommerfeld's half-plane problem, describing diffraction of a monochromatic wave by a semi-infinite perfect electric conducting (PEC) infinitesimally thin plate, is one of the cornerstones of classical electromagnetic scattering theory. Here we study its temporal analogue: a semi-infinite medium with the Drude dispersion, adjacent to air, whose plasma frequency is rapidly reduced to zero, thus turning it into air, while a monochromatic wave with frequency below the plasma frequency illuminates it under normal incidence. Unlike the PEC plate, the Drude medium supports evanescent penetration of the field prior to its temporal transition; the temporal interface here acts as the analogue of the sharp spatial boundary in the Sommerfeld problem, but now separates two time intervals rather than two spatial half-planes. We show that this transition forces the standing wave in air in front of the Drude medium and the evanescent fields inside it to propagate in both directions, while preserving their spatial field structure. In addition, we observe the excitation of additional electric and magnetic field waves generated by a "DC" current density associated with the time interface. Our findings offer insight into the electromagnetic wave behavior at temporal transitions in dispersive media and may find applications in time- and space-time metamaterials.

\end{abstract}

\maketitle

\section{\label{sec:introduction}Introduction}

In recent years, the study of wave interaction with four-dimensional (4D) material platforms, in which material parameters change with time while waves are present in the media, has opened new chapters in the manipulation of electromagnetic waves as they interact with matter \cite{caloz2019spacetime,engheta2020metamaterials,galiffi2022photonics,yin2022floquet,engheta2023four} offering a diverse set of applications, including antireflection temporal coating \cite{pacheco2020antireflection}, temporal aiming \cite{pacheco2020temporal}, time reversal and holography \cite{bacot2016time,moussa2023observation,jones2024time}, pulse shaping and reversal \cite{galiffi2023broadband,stefanini2026pulse-reveral}, nonreciprocity \cite{sounas2017non,li2022nonreciprocity}, and exotic energy exchanges between waves and matter \cite{li2021temporal,pacheco2025holding,ptitcyn2026photonic,galiffi2025electrodynamics,stefanini2025theory,castaldi2026analytic}.

In our previous work \cite{pacheco2025holding}, we studied how a propagating wave in an unbounded isotropic dielectric medium undergoes changes when the relative permittivity of the entire medium is rapidly changed from its original positive value to a nondispersive (non-Foster) negative value. Here, we explore the reverse phenomenon: a temporal change of the negative permittivity of a semi-infinite Drude medium to a positive value, and how such a time interface affects the features of a one-dimensional propagating wave. We assume the original negative-permittivity medium to be dispersive, following the Drude model, with its plasma frequency rapidly switched in time from a value larger than the signal frequency to a near-zero value, in essence, the medium is rapidly changed from a Drude metal to a simple dielectric (e.g., "air").

Sommerfeld's half-plane problem \cite{born2013principles} stands as one of the important diffraction problems in electromagnetic theory, addressing the intricate interaction of electromagnetic fields with the perfect electric conducting (PEC), infinitesimally thin boundary. It addresses the diffraction properties of the electromagnetic waves in the vicinity of PEC "knife edge", situated at the reference half-plane defined by $y=0$, $-\infty<x<+\infty$, and $-\infty<z<0$, as shown in the upper panel of Fig. \ref{fig:geo}.  In the present problem, we study the temporal analog of this canonical scattering configuration, in which the second medium is a metal (not necessarily a PEC, but more generally modeled with the Drude dispersion) for $-\infty<t<0$, and it becomes "air" for $0<t<+\infty$.  By anology, in our case here, we establish the interface at $t=0$, necessitating rigorous adherence to causality. Furthermore, our investigation employs a dispersive metal to characterize the half-space medium, deviating from the conventional use of PEC prevalent in Sommerfeld's analyses.

Related time-interface problems have been studied in half-space dispersive structures: a positive-to-positive permittivity transition has been shown to generate new frequencies \cite{fante1971transmission}, while a transition from air to an ionized plasma produces a time-independent magnetic field alongside frequency up-conversion \cite{wilks1988frequency}. Here we instead consider the reverse transition, from a dispersive negative permittivity (Drude) medium to air in a half-space geometry.

\section{\label{sec:geometry}Geometry and Statement of the Problem}

\begin{figure}[!t]
    \centering
    \includegraphics[width=\columnwidth]{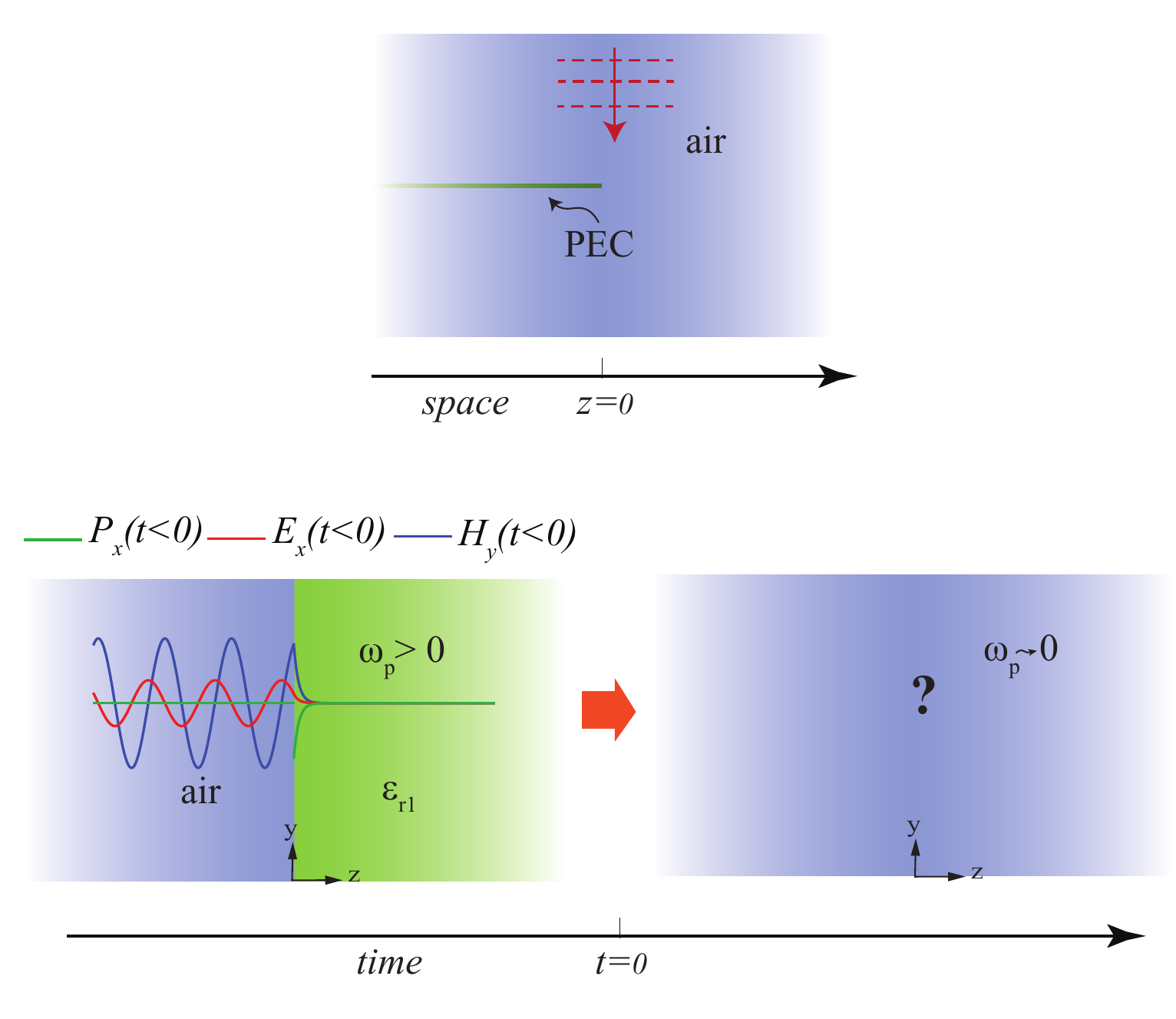}
    \caption{Upper panel: Sommerfeld's well known half-plane problem. A plane wave impinges on and is scattered by the half-plane perfect electric conducting (PEC) plate. Lower panel: A semi-infinite Drude material before ($t<0$) and after switch ($t>0$). The plasma frequency of the Drude medium on the right side of the $z$-axis undergoes a transition at $t=0$ from $\omega_p>0$ to $\omega_p \approx0$. }
    \label{fig:geo}
\end{figure}

Figure \ref{fig:geo} (lower panel) presents the geometry of our problem: the semi-infinite medium on the left ($z<0$) is air, and the medium on the right ($z>0$) exhibits negative relative permittivity $\varepsilon_{r1}$ for a monochromatic wave whose frequency is less than the plasma frequency of the medium.  At $t=0$, the medium on the right undergoes a transition in which its plasma frequency is rapidly changed to a very small value near zero. Although the medium for the $z>0$ region before $t=0$ represents a medium with Drude dispersion, for the sake of mathematical generality in our analysis we assume a Lorentzian dispersion with a very small resonance frequency, thus resembling the Drude dispersion, while keeping our analysis more general. We assume that the plasma frequency, $\omega_p$, is larger than the impinging wave frequency ($\omega$) before $t=0$ [such that $\varepsilon_{r1}(t<0)=-5$] and it becomes vanishingly small after the switch, such that $\varepsilon_r(t>0)\approx1$. We investigate this structure both analytically, using the Laplace transformation, and numerically using the Finite-Difference Time-Domain (FDTD) simulations, obtaining consistent and congruent results. In the following, we explain each approach and compare its respective results. 

Assuming an $x$-polarized monochromatic plane wave with frequency $\omega$ propagating normally to the boundary between the two media, in the steady-state pre-transition scenario (i.e., before $t=0$), the electric and magnetic field distributions along the $z-$axis can be presented by $E_{xi}^{t^-}$ and $H_{yi}^{t^-}$ (subscripts $i=r,l$ represent the right and left side of $z=0$, respectively, and superscript $t^-$ indicates times before $t=0$) as

\begin{subequations}
\label{eq:exr1}
\begin{align}
E_{x r}^{t^-} &= E_0 e^{-\alpha_r z} \cos(\omega t - \varphi), \label{eq:exr1a} \\
H_{y r}^{t^-} &= \frac{E_0}{c \mu_0}\frac{\alpha_r}{k_0} e^{-\alpha_r z} \sin(\omega t - \varphi) \label{eq:exr1b}
\end{align}
\end{subequations}

\begin{subequations}
\label{eq:exl1}
\begin{align}
 E_{x l}^{t^-} &= E_0\left[
 \cos(k_0 z) - \frac{\alpha_r}{k_0} \sin(k_0 z)
 \right] \cos(\omega t - \varphi), \label{eq:exl1a} \\
 H_{y l}^{t^-} &= \frac{E_0}{c \mu_0} \left[
 \sin(k_0 z) + \frac{\alpha_r}{k_0} \cos(k_0 z)
 \right] \sin(\omega t - \varphi) \label{eq:exl1b}
\end{align}
\end{subequations}

\noindent where $\alpha_r\equiv k_0 \sqrt{-\varepsilon_{r1}}$, $\varphi$, $k_0\equiv\omega/c$, and $c$ are the attenuation constant of the field on the right, the initial phase of the electric field at $t=0$, the wavenumber and the speed of light in vacuum, respectively. $E_0$ is the amplitude of the total electric field at $z=0$. 

The response of the material occupying the $z>0$ region is described by the Lorentzian model (with very small resonance frequency as mentioned above), derived from the following second-order differential equation governing the relationship between the electric polarization density($P_x$) and the electric field($E_x$),
 \begin{gather}
 \frac{d^2P_x}{dt^2}+\gamma\frac{dP_x}{dt}+\omega_0^2P_x=\varepsilon_0 \omega_p^2E_x,
 \label{eq:lorT}
  \end{gather}

\noindent where $\omega_p$, $\varepsilon_0$, $\gamma$, and $\omega_0$ are the plasma frequency, vacuum permittivity, damping, and resonance frequency.

To determine the electromagnetic fields after $t=0$, we consider Maxwell's equations in the Laplace domain and use the Laplace transform \cite{stratton2007electromagnetic} of $E_x (t, z)$ and $H_y(t, z)$ with respect to $t$, using the relations $\tilde{E}_x(s, z)=\int_0^{\infty} E_x(t, z) e^{-s t} dt$ and $\tilde{H}_y(s, z)=\int_0^{\infty} H_y(t, z) e^{-s t} d t$. Consequently, we obtain the following differential equation for $\tilde{E}_x(s, z)$:

\begin{equation}
\frac{\partial^2 E_x(z, s)}{\partial z^2}-\frac{s^2}{c^2}\left(1+\frac{\omega_p^2}{s^2+\omega_{0}^2
+\gamma s}\right) E_x(z, s)=\psi(z, s).
    \label{eq:diff}
\end{equation}

 \noindent where the inhomogeneous term on the right-hand side $\psi(z, s)\equiv-\mu_0\varepsilon_0 \left[sE_x(0^-,z)+\frac{\partial E_x(0^-,z)}{\partial t}\right]-\mu_0 \left[sP(0^-)+\frac{dP_x(0^-)}{dt}\right]+s^2\mu_0\frac{(s+\gamma)P_x(0^-)+\frac{dP_x(0^-)}{dt}}{s^2+\omega_0^2+\gamma s}$, comprising the initial values of the fields right after the moment of transition. It is worth noting that polarization ($P$), electric field ($E$), magnetic field ($H$), and the time derivatives $\partial E/\partial t$, $dP/dt$ are all continuous at $t=0$ in the entire space, which can be proved by the initial value theorem \cite{kalluri2018electromagnetics,solis2021time} or the balance of distributions \cite{gratus2021temporal}. 

\section{\label{sec:results}Results}

The solutions of Eq.~(\ref{eq:diff}) for both regions are obtained analytically as \cite{stratton2007electromagnetic},

\begin{equation}
 \begin{aligned}
  \tilde{E}_x(s, z>0)&=B_r e^{-K_r z}+\frac{e^{K_r z}}{2 K_r} \int e^{-K_r z} \psi(z, s) d z \\
&-\frac{e^{-K_r z}}{2 K_r} \int e^{K_r z} \psi(z, s) dz,  
    \label{eq:solr}
\end{aligned}
\end{equation}

\begin{equation}
 \begin{aligned}
\tilde{E}_x(s, z<0)&= A_l e^{K_l z}+\frac{e^{K_l z}}{2 K_l} \int e^{-K_l z} \psi(z, s) d z \\
&-\frac{e^{-K_l z}}{2 K_l} \int e^{K_l z} \psi(z, s) dz, 
    \label{eq:soll}
\end{aligned}
\end{equation}

\noindent where $K_r\equiv\frac{s^2}{c^2}\left(1+\frac{\omega_p^2}{s^2+\omega_{0}^2 +\gamma s}\right)$
and $K_l\equiv\frac{s}{c}$. $B_r$ and $A_l$ can be determined using the continuity of the electric and magnetic fields at $z=0$.
We can then find the expressions of the electric, magnetic, and polarization density fields in the time domain, utilizing the inverse Laplace transform. In the following, we adopt the notation of retarded time $\tilde{t}_{\rm ret}=t-z/c$ and advanced time $\tilde{t}_{\rm adv}=t+z/c$, associated with wave propagation in the $\pm z$ direction, respectively. Then the electric field after the time switching for $z>0$ reads

\begin{equation}
\begin{aligned}
E_{x r}^{t^{+}}/E_0 &= \frac{1}{2} e^{-\alpha_r c \tilde{t}_{\rm adv}} \cos\varphi
\\
&\quad + \left(\frac{1}{2} e^{\alpha_r c \tilde{t}_{\rm ret}} \cos\varphi \right) U(-\tilde{t}_{\rm ret})\\
&\quad - \frac{k_0}{2 \alpha_r} e^{-\alpha_r c \tilde{t}_{\rm adv}} \sin\varphi
\\
&\quad + \left( \frac{k_0}{2 \alpha_r} e^{\alpha_r c \tilde{t}_{\rm ret}} \sin\varphi \right) U(-\tilde{t}_{\rm ret})  \\
 &\quad + \frac{1}{2} \left[ \cos(\omega \tilde{t}_{\rm ret} - \varphi) + \frac{\alpha_r}{k_0} \sin(\omega \tilde{t}_{\rm ret} - \varphi) \right]\\ &\qquad \times U(\tilde{t}_{\rm ret}) \\
&\quad + \left[\frac{k_0}{2 \alpha_r} \left(1 + \frac{\alpha_r^2}{k_0^2}\right) \sin\varphi \right] U(\tilde{t}_{\rm ret}).
\end{aligned}
\label{eq:solrt}
\end{equation}

After $t=0$, the medium on the right-hand side becomes air. Consequently, the evanescent fields originally residing on the right, represented by the exponential term in Eq.~(\ref{eq:exr1a}), splits into two counter-propagating portions as expected in the presence of a time interface. These two components correspond to the first and second terms in Eq.~(\ref{eq:solrt}), respectively. However, the portion that propagates to the right (the second term) extends only down to \( z \geq ct \), constrained by the step function $U(-\tilde{t}_{\rm ret})$.
The third and fourth terms describe the electric field induced by the magnetic field, which was originally an evanescent field exponentially decaying (before $t=0$), but splits into two counter-propagating portions propagating in both directions upon switching. It is important to note that, prior to the switch, the electric and magnetic evanescent fields on the right side had a 90$^\circ$ phase difference, dictated by the negative permittivity. After \( t = 0 \), the permittivity becomes positive (as in air). As a result, the previous electric field begins to evolve dynamically, generating its own magnetic field in air. Analogously, the same happens for the previous magnetic field, which, after $t=0$, starts to propagate accompanied by its own new electric field. These effects are precisely captured by the third and fourth terms. An analogous mechanism can be found in Sommerfeld's half-plane problem, where the material mismatch at the edge generates a diffracted wave that radiates from it \cite{Keller1962Geometrical}, alongside the spatially reflected fields.

The fifth term in the solution of our problem Eq. (7) corresponds to the impinging wave from the air side, which was contributing to the standing wave described by Eq.~(\ref{eq:exl1a}) before $t=0$. Before the switch, it could not propagate into the region $z>0$ due to total reflection ($|R|=1$), but there was an evenescent field in the right region. Once the permittivity becomes positive on the right side, this impinging wave can now propagate up to $z \leq ct$, denoted by $U(\tilde{t}_{\rm ret})$. Notice that the left-propagating, originally reflected, wave from the interface before $t=0$   does not appear here, as we are analyzing the field on the right side for now. For similar reasons, the exponential terms in the first four terms in Eq.~(\ref{eq:solrt}) remain finite because $z>0$ limits the exponential growth in space associated with the advanced time $\tilde{t}_{\rm adv}=t+z/c$, while the step function $U(-\tilde{t}_{\rm ret})$, as we explained, truncates the exponential growth in time associated with the retarded time $\tilde{t}_{\rm ret}=t-z/c$.

More interestingly, the last term in Eq.~(\ref{eq:solrt}) arises immediately after setting \( \omega_p \approx 0 \), corresponding to a sudden disappearance of the plasma response (which can model a sudden large increase of the mass of free charged particles). At this moment, the free charged particles that were previously oscillating driven by the monochromatic electric field of the signal, responsible for the negative permittivity of the material, begin to move freely with constant velocity determined at the transition time, not affected by the electric field because they have a large mass now, giving rise to a linearly increasing polarization density field $P_x(z,t)=E_0e^{-\alpha_r z}\varepsilon_0(\varepsilon_{r1}-1)(\cos\varphi+ \omega t \sin\varphi) U(z)$. Taking the time derivative  ($dP_x/dt$) of this polarization results in a DC polarization current density $J_{px}(z,t)=E_0e^{-\alpha_r z}\omega\varepsilon_0(\varepsilon_{r1}-1)(\sin\varphi) U(z)$, whose amplitude decays exponentially in space. This position-dependent DC current density, can be seen as the temporal analogue of the excitation of a Wiggler mode \cite{jiang1975wave,wilks1988frequency,kalluri1988reflection}, generating a burst of electric field, which corresponds to the last term in Eq.~(\ref{eq:solrt}).

The corresponding magnetic field for $z>0$ is given by
 
\begin{equation}
\begin{aligned}
H_{y r}^{t^{+}}/\frac{E_0}{c \mu_0} &=  -\frac{1}{2} e^{-\alpha_r c \tilde{t}_{\rm adv}} \cos\varphi\\
&\quad
+ \left( \frac{1}{2} e^{\alpha_r c \tilde{t}_{\rm ret}} \cos\varphi \right) U(-\tilde{t}_{\rm ret})  \\
&\quad + \frac{k_0}{2 \alpha_r} e^{-\alpha_r c \tilde{t}_{\rm adv}} \sin\varphi\\
&\quad + \left(\frac{k_0}{2\alpha_r} e^{\alpha_r c \tilde{t}_{\rm ret}} \sin\varphi \right) U(-\tilde{t}_{\rm ret}) \\
&\quad + \frac{1}{2} \left[ \cos(\omega \tilde{t}_{\rm ret} - \varphi) + \frac{\alpha_r}{k_0} \sin(\omega \tilde{t}_{\rm ret} - \varphi) \right] \\ &\qquad \times U(\tilde{t}_{\rm ret}) \\
&\quad - \frac{k_0}{2 \alpha_r} \left(1 + \frac{\alpha_r^2}{k_0^2} \right)  \left[2 e^{-\alpha_r z} - U(\tilde{t}_{\rm ret})\right]\sin\varphi.
\end{aligned}
\label{eq:solrth}
\end{equation}

The terms shown in Eq. (8) are the corresponding magnetic fields related to the electric fields described in Eq. (7).   
The time-domain expression for the electric field $E_{xl}^{t^+}$ in the semi-space $z < 0$, categorized in the same manner and grouped into five terms are given by
\begin{equation}
\begin{aligned}
E_{x l}^{t^{+}}/E_0 &= 
 \left[ \frac{1}{2} \cos(\omega \tilde{t}_{\rm ret} - \varphi) + \frac{\alpha_r}{2 k_0} \sin(\omega \tilde{t}_{\rm ret} - \varphi) \right] \\
 &\quad + \left[ \frac{1}{2} \cos(\omega \tilde{t}_{\rm adv} - \varphi) - \frac{\alpha_r}{2 k_0} \sin(\omega \tilde{t}_{\rm adv}- \varphi) \right] \\ &\qquad \times U(-\tilde{t}_{\rm adv}) \\
&\quad + \left(\frac{1}{2} e^{-\alpha_r c \tilde{t}_{\rm adv}} \cos\varphi \right) U(\tilde{t}_{\rm adv})\\
&\quad - \left(\frac{k_0}{2 \alpha_r} e^{-\alpha_r c \tilde{t}_{\rm adv}} \sin\varphi \right)  U(\tilde{t}_{\rm adv})\\
&\quad + \left[\frac{k_0}{2 \alpha_r} \left(1 + \frac{\alpha_r^2}{k_0^2} \right) \sin\varphi \right] U(\tilde{t}_{\rm adv}).
\end{aligned}
\label{eq:sollt}
\end{equation}

\noindent Similar to Eq.(\ref{eq:solrt}), the first two terms represent two counter-propagating waves arising from the standing wave that existed in front of the interface before \( t = 0 \), which becomes free to propagate in both directions after \( t = 0 \) once the spatial interface is removed by switching the plasma frequency. As before, the left-going wave includes a step function. The third and fourth terms describe the propagation of evanescent tails, which originate from the electric and magnetic fields on the right-hand side and now propagate into the left region. The fifth term corresponds to a constant electric field propagating to the left. This field is generated by the DC current density induced on the right side, as previously discussed. This observation is analogous to the behavior observed in the evolution of the electric field for \( z > 0 \).

Similarly, the corresponding magnetic field for $z<0$ is given by
 
\begin{equation}
\begin{aligned}
H_{y l}^{t^{+}}/\frac{E_0}{c \mu_0} &= 
\frac{1}{2} \left[ \cos(\omega \tilde{t}_{\rm ret} - \varphi) + \frac{\alpha_r}{k_0} \sin(\omega \tilde{t}_{\rm ret} - \varphi) \right] \\
&\quad - \frac{1}{2} \left[ \cos(\omega \tilde{t}_{\rm adv} - \varphi) - \frac{\alpha_r}{k_0} \sin(\omega \tilde{t}_{\rm adv} - \varphi) \right] \\ &\qquad \times U(-\tilde{t}_{\rm adv})\\
&\quad - \left(\frac{1}{2} e^{-\alpha_r c \tilde{t}_{\rm adv}} \cos\varphi \right) U(\tilde{t}_{\rm adv})\\
&\quad + \left(\frac{k_0}{2\alpha_r} e^{-\alpha_r c \tilde{t}_{\rm adv}} \sin\varphi \right)U(\tilde{t}_{\rm adv}) \\
&\quad - \left[\frac{k_0}{2\alpha_r} \left(1 + \frac{\alpha_r^2}{k_0^2} \right) \sin\varphi \right] U(\tilde{t}_{\rm adv}).
\end{aligned}
\label{eq:sollth}
\end{equation}

\begin{figure}
    \centering
    \includegraphics[width=\columnwidth]{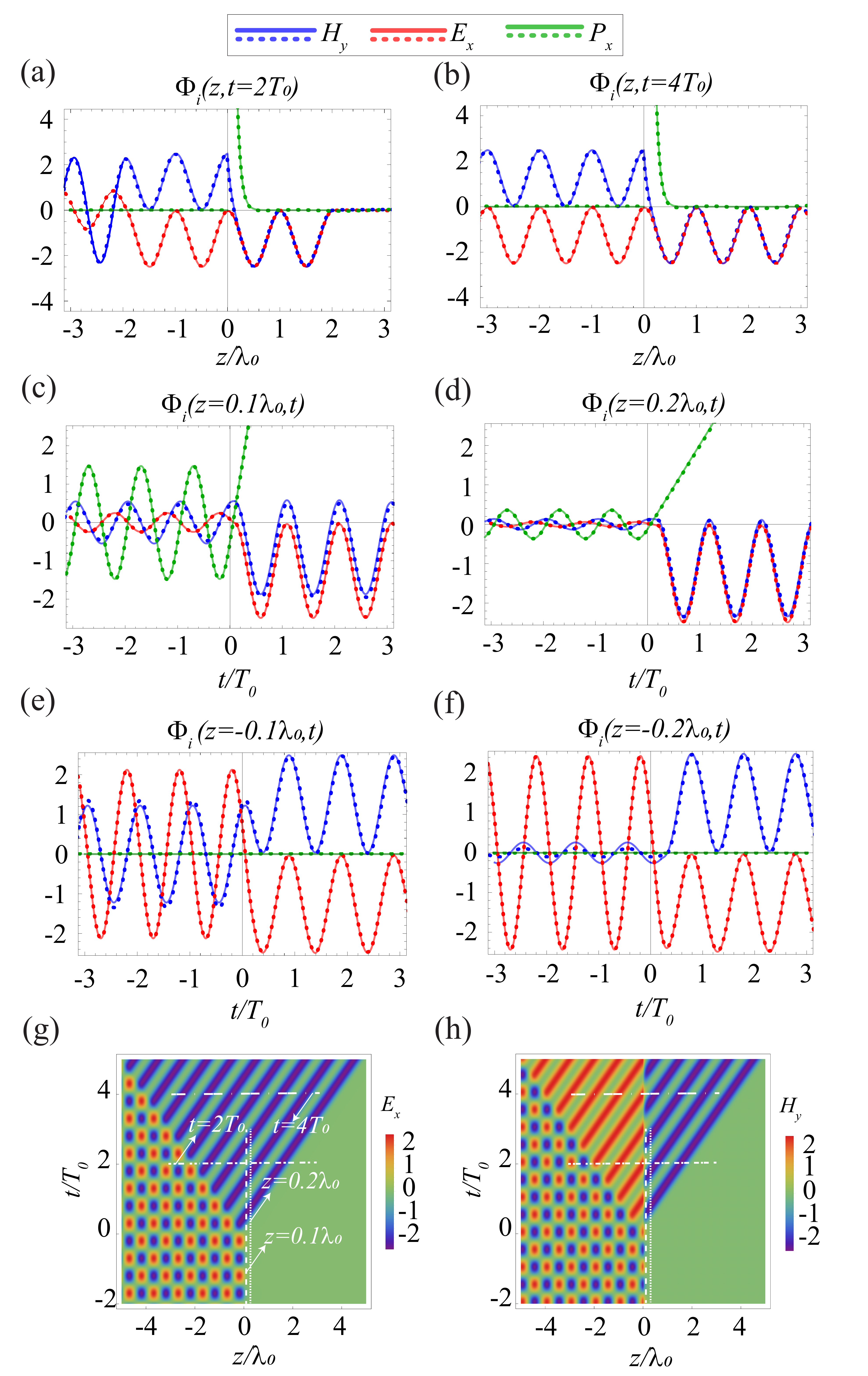}
    \caption{A snapshot at (a) $t=2T_0$ (b) $t=4T_0$ of the electric, magnetic, and polarization density fields versus space.  Time evolution of these fields at (c) $z=0.1 \lambda_0$ (d) $z=0.2 \lambda_0$ (e) $z=-0.1 \lambda_0$ (f) $z=-0.2 \lambda_0$. The dotted curves are the results of an FDTD numerical simulation. $\Phi_i(z,t)$ represents $E_x$, $H_y$, and $P_x$ at spatial and temporal points of $z$ and $t$ , respectively. Space and time evolution of  (g) $x-$component of the electric and (h) $y-$component of the magnetic field. White dashed lines denote the positions and instances used to plot the fields against space and time in panels (a)-(d). The fields are shown in normalized units with the speed of light set to unity ($c=1$)}
    \label{fig:fig2}
\end{figure}

The physics at hand becomes clearer when visualized through their temporal and spatial distribution plots. Figures \ref{fig:fig2} (a-b) illustrate the spatial distribution of the electric, magnetic, and polarization density fields at two moments in time; $t=2T_0$, and $t=4T_0$ after the temporal switch at $t=0$, respectively, delineated across both regions. Throughout this study the plasma and resonance frequency of the right-side medium after the transition are assumed to be $\omega_{p}^+=10^{-6}\omega$ and $\omega_0^+=0.1 \omega_p^+$, respectively. 

In Fig.~\ref{fig:fig2}(c-d), the time evolution of $E_x$, $H_y$, and $P_x$ is depicted at two locations in space on the right side, $z=0.1\lambda_0$ and $z=0.2\lambda_0$, respectively. Notably, as the medium undergoes the transition at $t=0$, the standing wave originating from the left side penetrates and a portion propagates to the right side, resulting in a noticeable increase in the amplitudes of both the electric and magnetic fields. Furthermore, the presence of DC values for the electric and magnetic fields is discernible in these panels. Additionally, the behavior of the $x$-component of the polarization density field after the switch is interesting, as evidenced by the growth of $P_x$ proportional to $t$ (due to the motion of heavy charged particles with constant speed after $t=0$, which results in the time-invariant current density in that medium ($J_x=dP_x/dt$).
Fig.~\ref{fig:fig2}(e-f), show the time evolution of the same field components at two locations in space on the left side, $z=-0.1\lambda_0$ and $z=-0.2\lambda_0$. DC components of the electric and magnetic fields are also observed on the left side. A comparison with the earlier panels (c–d) reveals a clear polarity difference in the magnetic field on the two sides. 

The results obtained from FDTD simulations are shown in Fig. \ref{fig:fig2}(a-f) in dashed curves, demonstrating excellent agreement with our analytical study.

\section{\label{sec:discussion}Discussion}

Several key observations can be deduced from Fig. \ref{fig:fig2}. We notice that the pre-transition standing waves begin to propagate in both directions after $t=0$. In addition, the polarization density field experiences an exponential decrease in space, characterized by the factor $e^{-\alpha_r z}$. Both the electric and magnetic fields exhibit DC components, albeit with distinct polarities on opposite sides of the medium. This contrast in polarity in the DC component of the magnetic field arises from the presence of the time-invariant current density generated during the temporal change in the region $z>0$ close to the boundary $z=0$, whose manifestation can be observed in the variation of the spatial derivative of the magnetic field near this boundary. In order to gain some insight into this result, let us consider how conceptually one can achieve the temporal transition of  $\omega_p$ from a non-zero value to a value close to zero.  Since the plasma frequency is inversely proportional to the mass of the free charged particle, as discussed in the previous section, we can model this transition in a Drude medium as a sudden large increase in the effective mass of the charged particles at the transition time $t=0$, while keeping the velocity and charge of the particle intact, thus bringing the plasma frequency to zero.  To achieve such a transition, we require a large amount of energy.  With this hypothetical assumption, immediately after $t=0$, the charged particles continue with their motion at their velocity determined at $t=0$.  Since their mass is now very large, they do not slow down due to presence of the electric field, thus creating a DC current density (and a linearly increasing polarization density with time) that generates the DC component of the electric and magnetic fields propagating in the two regions shown in our results after $t=0$.  In our scenario, with $\varepsilon_{r1}=-5$, the evanescent behavior is particularly sharp, leading to a limited spatial extent for this DC current density. Consequently, this causes the DC magnetic field to exhibit opposite signs in the right and left regions. It is worth mentioning that such electric and magnetic fields that are generated by this DC current and propagating to the left and right regions should get their energy from slowing down of such moving free charged particles with large mass.  However, since these particles are assumed to have a large mass after the temporal transition, they would take a long time to slow down.  In principle, in the limit of infinite mass and zero loss/collision, they will not slow down.  But this implies that in this limit, we need to exert "infinite" energy from outside to make the mass of such charged particles large at $t=0$, and that is where the energy of electric and magnetic fields of the DC current comes from.  

\section{\label{sec:summary}Summary}

In this work, we have theoretically studied how an abrupt temporal reduction of the plasma frequency of a semi-infinite Drude material to near zero frequency impacts a normally incident wave. We have provided analytical results for the electric and magnetic fields in all regions before and after this temporal interface, and have compared them with FDTD numerical simulations, confirming good agreement.  The results reveal interesting wave dynamics in this time-varying scenario, which is effectively the temporal analogue of the Sommerfeld half-plane problem, and offers insight into the wave physics of temporal and spatiotemporal metamaterials. 

\begin{acknowledgments}
We acknowledge the support from the Simons Foundation/Collaboration on Symmetry-Driven Extreme Wave Phenomena (grant $\# $SFI-MPS-EWP-00008530-04).
\end{acknowledgments}

\nocite{*}
\bibliography{sorsamp}%

@article{gratus2021temporal,
  title={Temporal boundaries in electromagnetic materials},
  author={Gratus, Jonathan and Seviour, Rebecca and Kinsler, Paul and Jaroszynski, Dino A},
  journal={New Journal of Physics},
  volume={23},
  number={8},
  pages={083032},
  year={2021},
  publisher={IOP Publishing}
}

@book{kalluri2018electromagnetics,
  title={Electromagnetics of time varying complex media: frequency and polarization transformer},
  author={Kalluri, Dikshitulu K},
  year={2018},
  publisher={CRC Press}
}

@article{li2022nonreciprocity,
  title={Nonreciprocity and Faraday rotation at time interfaces},
  author={Li, Huanan and Yin, Shixiong and Al{\`u}, Andrea},
  journal={Physical Review Letters},
  volume={128},
  number={17},
  pages={173901},
  year={2022},
  publisher={APS}
}

@article{jones2024time,
  title={Time-reflection of microwaves by a fast optically-controlled time-boundary},
  author={Jones, Thomas R and Kildishev, Alexander V and Segev, Mordechai and Peroulis, Dimitrios},
  journal={Nature Communications},
  volume={15},
  number={1},
  pages={6786},
  year={2024},
  publisher={Nature Publishing Group UK London}
}

@article{moussa2023observation,
  title={Observation of temporal reflection and broadband frequency translation at photonic time interfaces},
  author={Moussa, Hady and Xu, Gengyu and Yin, Shixiong and Galiffi, Emanuele and Ra’di, Younes and Al{\`u}, Andrea},
  journal={Nature Physics},
  volume={19},
  number={6},
  pages={863--868},
  year={2023},
  publisher={Nature Publishing Group UK London}
}

@article{pacheco2025holding,
  title={Holding and amplifying electromagnetic waves with temporal non-Foster metastructures},
  author={Pacheco-Pe{\~n}a, Victor and Kiasat, Yasaman and Sol{\'\i}s, Diego M and Edwards, Brian and Engheta, Nader},
  journal={Nature Communications},
  volume={16},
  number={1},
  pages={1--11},
  year={2025},
  publisher={Nature Publishing Group}
}

@article{yin2022floquet,
  title={Floquet metamaterials},
  author={Yin, Shixiong and Galiffi, Emanuele and Al{\`u}, Andrea},
  journal={ELight},
  volume={2},
  number={1},
  pages={8},
  year={2022},
  publisher={Springer}
}

@article{caloz2019spacetime,
  title={Spacetime metamaterials—Part II: Theory and applications},
  author={Caloz, Christophe and Deck-Leger, Zoe-Lise},
  journal={IEEE Transactions on Antennas and Propagation},
  volume={68},
  number={3},
  pages={1583--1598},
  year={2019},
  publisher={IEEE}
}

@article{galiffi2022photonics,
  title={Photonics of time-varying media},
  author={Galiffi, Emanuele and Tirole, Romain and Yin, Shixiong and Li, Huanan and Vezzoli, Stefano and Huidobro, Paloma A and Silveirinha, M{\'a}rio G and Sapienza, Riccardo and Al{\`u}, Andrea and Pendry, John B},
  journal={Advanced Photonics},
  volume={4},
  number={1},
  pages={014002--014002},
  year={2022},
  publisher={Society of Photo-Optical Instrumentation Engineers}
}

@article{engheta2020metamaterials,
  title={Metamaterials with high degrees of freedom: space, time, and more},
  author={Engheta, Nader},
  journal={Nanophotonics},
  volume={10},
  number={1},
  pages={639--642},
  year={2020},
  publisher={De Gruyter}
}

@article{wilks1988frequency,
  title={Frequency up-conversion of electromagnetic radiation with use of an overdense plasma},
  author={Wilks, SC and Dawson, JM and Mori, WB},
  journal={Physical review letters},
  volume={61},
  number={3},
  pages={337},
  year={1988},
  publisher={APS}
}

@article{solis2021time,
  title={Time-varying materials in the presence of dispersion: plane-wave propagation in a Lorentzian medium with temporal discontinuity},
  author={Sol{\'\i}s, Diego M and Kastner, Raphael and Engheta, Nader},
  journal={Photonics Research},
  volume={9},
  number={9},
  pages={1842--1853},
  year={2021},
  publisher={Optica Publishing Group}
}

@article{engheta2023four,
  title={Four-dimensional optics using time-varying metamaterials},
  author={Engheta, Nader},
  journal={Science},
  volume={379},
  number={6638},
  pages={1190--1191},
  year={2023},
  publisher={American Association for the Advancement of Science}
}

@article{bacot2016time,
  title={Time reversal and holography with spacetime transformations},
  author={Bacot, Vincent and Labousse, Matthieu and Eddi, Antonin and Fink, Mathias and Fort, Emmanuel},
  journal={Nature Physics},
  volume={12},
  number={10},
  pages={972--977},
  year={2016},
  publisher={Nature Publishing Group UK London}
}

@article{galiffi2023broadband,
  title={Broadband coherent wave control through photonic collisions at time interfaces},
  author={Galiffi, Emanuele and Xu, Gengyu and Yin, Shixiong and Moussa, Hady and Ra’di, Younes and Al{\`u}, Andrea},
  journal={Nature Physics},
  volume={19},
  number={11},
  pages={1703--1708},
  year={2023},
  publisher={Nature Publishing Group UK London}
}

@article{sounas2017non,
  title={Non-reciprocal photonics based on time modulation},
  author={Sounas, Dimitrios L and Al{\`u}, Andrea},
  journal={Nature Photonics},
  volume={11},
  number={12},
  pages={774--783},
  year={2017},
  publisher={Nature Publishing Group UK London}
}

@article{pacheco2020antireflection,
  title={Antireflection temporal coatings},
  author={Pacheco-Pe{\~n}a, Victor and Engheta, Nader},
  journal={Optica},
  volume={7},
  number={4},
  pages={323--331},
  year={2020},
  publisher={Optica Publishing Group}
}

@article{li2021temporal,
  title={Temporal parity-time symmetry for extreme energy transformations},
  author={Li, Huanan and Yin, Shixiong and Galiffi, Emanuele and Al{\`u}, Andrea},
  journal={Physical Review Letters},
  volume={127},
  number={15},
  pages={153903},
  year={2021},
  publisher={APS}
}

@article{pacheco2020temporal,
  title={Temporal aiming},
  author={Pacheco-Pe{\~n}a, Victor and Engheta, Nader},
  journal={Light: Science \& Applications},
  volume={9},
  number={1},
  pages={129},
  year={2020},
  publisher={Nature Publishing Group UK London}
}

@book{born2013principles,
  title={Principles of optics: electromagnetic theory of propagation, interference and diffraction of light},
  author={Born, Max and Wolf, Emil},
  year={2013},
  publisher={Elsevier}
}

@book{stratton2007electromagnetic,
  title={Electromagnetic theory},
  author={Stratton, Julius Adams},
  year={2007},
  publisher={John Wiley \& Sons}
}

@article{kalluri1988reflection,
  title={On reflection from a suddenly created plasma half-space: Transient solution},
  author={Kalluri, Dikshitulu K},
  journal={IEEE transactions on plasma science},
  volume={16},
  number={1},
  pages={11--16},
  year={1988},
  publisher={IEEE}
}

@article{jiang1975wave,
  title={Wave propagation and dipole radiation in a suddenly created plasma},
  author={Jiang, Ching-Lin},
  journal={IEEE Transactions on Antennas and Propagation},
  volume={23},
  number={1},
  pages={83--90},
  year={1975},
  publisher={IEEE}
}

@article{fante1971transmission,
  title={Transmission of electromagnetic waves into time-varying media},
  author={Fante, R},
  journal={IEEE Transactions on Antennas and Propagation},
  volume={19},
  number={3},
  pages={417--424},
  year={1971},
  publisher={IEEE}
}

@article{ptitcyn2026photonic,
  title={Photonic temporal illusion},
  author={Ptitcyn, Grigorii and M. Sol{\'\i}s, D and Mirmoosa, MS and Engheta, Nader},
  journal={Optical Materials Express},
  volume={16},
  number={4},
  pages={874--890},
  year={2026},
  publisher={Optica Publishing Group}
}

@article{galiffi2025electrodynamics,
  title={Electrodynamics of photonic temporal interfaces},
  author={Galiffi, Emanuele and Sol{\'\i}s, Diego Mart{\'\i}nez and Yin, Shixiong and Engheta, Nader and Al{\`u}, Andrea},
  journal={Light: Science \& Applications},
  volume={14},
  number={1},
  pages={338},
  year={2025},
  publisher={Nature Publishing Group UK London}
}

@article{stefanini2025theory,
  title={Theory and experimental observation of scattering by a space-time corner},
  author={Stefanini, Luca and Galiffi, Emanuele and Yin, Shixiong and Singh, Sahitya and Sol{\'\i}s, Diego M and Engheta, Nader and Toscano, Alessandro and Ramaccia, Davide and Bilotti, Filiberto and Al{\`u}, Andrea},
  journal={Physical Review Letters},
  volume={135},
  number={11},
  pages={113802},
  year={2025},
  publisher={APS}
}

@article{castaldi2026analytic,
  title={Analytic Inverse Design of Temporal Metamaterials via Space-Time Duality},
  author={Castaldi, Giuseppe and Coppolaro, Marino and Moccia, Massimo and Rizza, Carlo and Engheta, Nader and Galdi, Vincenzo},
  journal={Physical Review Letters},
  volume={136},
  number={20},
  pages={206901},
  year={2026},
  publisher={APS}
}

@article{stefanini2026pulse-reveral,
  title = {Pulse reversal in adiabatically modulated plasmonic waveguides},
  author = {Stefanini, Luca and Khurgin, Jacob and Al\'u, Andrea},
  journal = {Phys. Rev. Appl.},
  volume = {25},
  issue = {4},
  pages = {L041002},
  numpages = {7},
  year = {2026},
  month = {Apr},
  publisher = {American Physical Society},
  doi = {10.1103/qm62-c6m6},
  url = {https://link.aps.org/doi/10.1103/qm62-c6m6}
}

@article{Keller1962Geometrical,
author = {Joseph B. Keller},
journal = {J. Opt. Soc. Am.},
number = {2},
pages = {116--130},
publisher = {Optica Publishing Group},
title = {Geometrical Theory of Diffraction$\ast$},
volume = {52},
month = {Feb},
year = {1962},
url = {https://opg.optica.org/abstract.cfm?URI=josa-52-2-116},
doi = {10.1364/JOSA.52.000116},
}

\end{document}